\documentclass[twocolumn,superscriptaddress,showpacs,preprintnumbers,amsmath,amssymb,prb]{revtex4}

\usepackage[dvipdfmx]{graphicx}
\usepackage{dcolumn}
\usepackage{bm}
\newcommand{\msr}{$\mu$SR}
\newcommand{\sro}{Sr$_2$IrO$_4$}
\newcommand{\bro}{Ba$_2$IrO$_4$}
\newcommand{\sros}{Sr$_3$Ir$_2$O$_7$}
\newcommand{\sroi}{SrIrO$_3$}
\newcommand{\srom}{Sr$_{n+1}$Ir$_{n}$O$_{3n+1}$}

\begin{document}

\preprint{APS}
\title{Evidence for ordered magnetic moments at oxygen sites in antiferromagnetic \sro\ and \sros}

\author{M. Miyazaki}
\affiliation{Muon Science Laboratory and Condensed Matter Research Center, Institute of Materials Structure Science, High Energy Accelerator Research Organization (KEK), Tsukuba, Ibaraki 305-0801, Japan}
\author{R. Kadono}\thanks{Corresponding author: ryosuke.kadono@kek.jp}
\affiliation{Muon Science Laboratory and Condensed Matter Research Center, Institute of Materials Structure Science, High Energy Accelerator Research Organization (KEK), Tsukuba, Ibaraki 305-0801, Japan}
\affiliation{Department of Materials Structure Science, Graduate University for Advanced Studies, Tsukuba, Ibaraki 305-0801, Japan}
\author{M. Hiraishi} 
\affiliation{Muon Science Laboratory and Condensed Matter Research Center, Institute of Materials Structure Science, High Energy Accelerator Research Organization (KEK), Tsukuba, Ibaraki 305-0801, Japan}
\author{A. Koda}
\affiliation{Muon Science Laboratory and Condensed Matter Research Center, Institute of Materials Structure Science, High Energy Accelerator Research Organization (KEK), Tsukuba, Ibaraki 305-0801, Japan}
\affiliation{Department of Materials Structure Science, Graduate University for Advanced Studies, Tsukuba, Ibaraki 305-0801, Japan}
\author{K.~M. Kojima}
\affiliation{Muon Science Laboratory and Condensed Matter Research Center, Institute of Materials Structure Science, High Energy Accelerator Research Organization (KEK), Tsukuba, Ibaraki 305-0801, Japan}
\affiliation{Department of Materials Structure Science, Graduate University for Advanced Studies, Tsukuba, Ibaraki 305-0801, Japan}
\author{K. Ohashi}
\affiliation{Department of Advanced Materials, Graduate School of Frontier Sciences, University of Tokyo, Kashiwa, Chiba 277-8561, Japan}
\author{T. Takayama}
\affiliation{Max Planck Institute for Solid State Research, Heisenbergstrasse 1, 70569 Stuttgart, Germany}
\author{H. Takagi}
\affiliation{Max Planck Institute for Solid State Research, Heisenbergstrasse 1, 70569 Stuttgart, Germany}
\affiliation{Department of Physics, Graduate School of Science, University of Tokyo, Tokyo 113-0033, Japan}

%

\begin{abstract}

In this study, the magnetic ground state of iridium perovskites (\srom, where $n=1$, 2, and $\infty$) was considered using muon spin spectroscopy (\msr). When probed by muons in \sro\ and \sros\ ($n=1,2$), the internal field ($B_{\rm loc}$) showed clear sign of a magnetic order in two stages at transition temperatures $T_{\rm N}\simeq230$ K and $T_{\rm m}\simeq90$ K in \sro\ and $T_{\rm N}\simeq280$ K and $T_{\rm m}\simeq70$ K in \sros, respectively. In contrast, no long-range magnetic order was observed in orthorhombic  \sroi\ ($n=\infty$).  Based on the known magnetic structure in \sro\ and \sros, we successfully identified muon sites in these compounds from the magnitude of $B_{\rm loc}$ in the first stage ($T_{\rm m}\le T\le T_{\rm N}$).  Below $T_{\rm m}$, $B_{\rm loc}$ probed by a fraction of muons occupying sites near the apical oxygen of IrO$_6$ octahedra exhibited a further increase but remained mostly unchanged for sites close to the in-plane oxygen.  While such behavior cannot be explained by the alteration of the Ir spin structure, it is consistent with the selective appearance of ordered magnetic moments on the apical oxygen.  
The oxygen polarization was also in line with the reported magnetization anomalies in these compounds below $\sim T_{\rm m}$. A possible link between the oxygen polarization and ferroelectric (multiferroic) behavior in \sro\ was considered.


\end{abstract}

\maketitle
\section{Introduction}
Layered iridium perovskites (Ruddlesden-Popper series: \srom, where $n=1,2,...$) serve as an intriguing playground for exploiting the novel properties of $d$ electrons arising from competition between the spin-orbit (SO) interaction, crystal field, and Coulomb interaction (electronic correlation) of comparable energy in $5d$ transition metal oxides. 
Recent studies with a variety of microscopic probes have established that the monolayer compound \sro\ is a Mott insulator realized within $t_{2g}$ manifolds that are reorganized by the SO interaction: the $t_{2g}$ band splits into two narrow subbands mainly composed of $J_{\rm eff}=3/2$ and $J_{\rm eff}=1/2$ spin-orbital states, where the latter is further split by the Coulomb interaction ($U$) into upper and lower Hubbard bands upon half-filling.\cite{Kim:08,Kim:09}  The bandwidth ($W$) increases with the number of IrO layers ($n$), which leads to the suppression of the Mott transition ($W>U$) and emergence of an anomalous metallic state in \sroi.\cite{Moon:08} 

Regarding magnetic properties, \sro\ is known to exhibit weak ferromagnetism below the Curie temperature $T_{\rm C}\simeq 230$ K in accordance with the Mott insulating phase, which is characterized by small saturation moments ($\mu_{\rm s}\sim0.1\mu_B$/Ir).\cite{Crawford:94,Shimura:95,Cao:98}  This is now understood to come from a canted antiferromagnetic (AF) order within the IrO$_2$ ($ab$) plane,\cite{Kim:08,Ye:13, Lovesey:12}  where the spin canting is induced by the strong SO interaction mapped onto the pseudo-dipolar interaction and the Dzyaloshinsky-Moriya interaction between the $J_{\rm eff}=1/2$ isospins.\cite{Jackeli:09}  While similar behavior has been reported for \sros\  ($\mu_{\rm s}<0.04\mu_B$/Ir),\cite{Cao:02} recent studies using resonant x-ray scattering have shown that the magnetic structure is a G-type AF\cite{Boseggia:12} with $c$-axis collinear moments,\cite{JWKim:12,Fujiyama:12} where the spin flop from \sro\ to \sros\ is attributed to the three-dimensional (cubic) character of the $J_{\rm eff}=1/2$ ground state that mediates the inter-layer pseudo-dipolar interaction common to the in-plane.\cite{JWKim:12}

However, these compounds exhibit complicated magnetic properties at low temperatures (and under weak magnetic fields) that do not necessarily fit into the current framework of effective theory. Recent studies on \sro\ have revealed a magnetic anomaly at low temperatures that is quite sensitive to an external field. The magnetization is anomalously reduced below $T_{\rm M}\sim 110$ K at 0.05 T (where $T_{\rm M}$ is reduced to $\sim$50 K at 0.15 T), which suggests an enhanced  antiferromagnetic correlation below $T_{\rm M}$.\cite{Chikara:09,Li:13}  The anomaly has also been shown to be closely linked with the charge sector;  the dielectric response exhibits crossover from relaxor-like to ferroelectric behavior as the temperature is decreased past $T_{\rm M}$. These features correlate with the reduction in the Ir-O-Ir bond angle, which suggests a change in spin canting induced by alteration of the exchange coupling.  While the situation is not as clear for \sros, earlier reports in the literature have reported the observation of a rapid decease in magnetization at $T_{\rm D}\simeq 50$ K and reversal in the basal plane below 20 K under field-cooled conditions at 0.01 T.  This has only been observed at low magnetic induction (below $\sim0.2$ T).\cite{Cao:02,Li:13}  Again, these anomalies exhibit a certain correlation with the charge sector as inferred from the sharp increase in the electrical resistivity.\cite{Chikara:09,Cao:02,Li:13}

In contrast to these two cases, no clear anomaly suggesting a phase transition has been observed in the bulk properties of orthorhombic \sroi\ ($n=\infty$) over the temperature range from ambient to the liquid helium temperature.   The electrical resistivity is metallic above $\sim$45 K, where it exhibits a minimum, and changes to localizing behavior at lower temperatures.\cite{Zhao:08} In addition, a positive magnetoresistance has been observed below $\sim$170 K.\cite{Zhao:08}  Through a band calculation considering the SO interaction and optical conductivity, the carrier has been inferred to have a large effective mass due to electronic correlation.\cite{Moon:08}  

Muon spin spectroscopy (\msr) is a unique probing technique for investigating the {\sl local} magnetic structure and spin dynamics with extreme sensitivity, which is useful for studying magnetism arising from Ir moments to which neutron diffraction techniques are difficult to apply because of the large neutron absorption cross-section of Ir nuclei.  In particular, once the muon site(s) in the relevant compound is confirmed, the local magnetic field probed by the muon serves as a true measure of the order parameter because it is directly proportional to the magnitude of local magnetic moments.  Note that the information gained from \msr\ is complementary to that from diffraction-based techniques (x-ray and neutron scattering), as the latter corresponds to the information averaged over an entire crystal volume. 

In this paper, we report \msr\ measurements on powder samples of \srom\ ($n=1,2$ and $\infty$) under a zero external field that establish the development of a long-range magnetic order  in \sro\ ($T_{\rm N}$ defined by ``onset" $\simeq 230$ K) and in \sros\ ($T_{\rm N}\simeq 280$ K).  More importantly, we demonstrate the presence of a secondary magnetic phase at lower temperatures (below $T_{\rm m}\simeq 90$ K in \sro\ and $T_{\rm m}\simeq 70$ K in \sros)  that can be discerned by the increase in spontaneous local fields at muon sites located near the apical oxygen of the IrO$_6$ octahedron.
This result provides a microscopic basis for understanding the magnetic anomalies observed in the bulk properties of these compounds below $T_{\rm m}$ and suggests that the order parameter of the secondary magnetic phase is not solely determined by the magnitude of Ir local moments. We show that the $\mu$SR result below $T_{\rm m}$ is consistently reproduced by postulating the presence of small ordered moments at the apical oxygen in those two compounds, which is in line with the above anomalies in their bulk properties. Meanwhile, the absence of magnetism for $n=\infty$ supports the scenario that the $t_{2g}$ band width is a monotonic function of the co-ordination number in this series of Ir perovskites.

\section{Experiment}
\subsection{Specimen}
\begin{figure}[b]
	\centering
	\includegraphics[width=0.75\linewidth,clip]{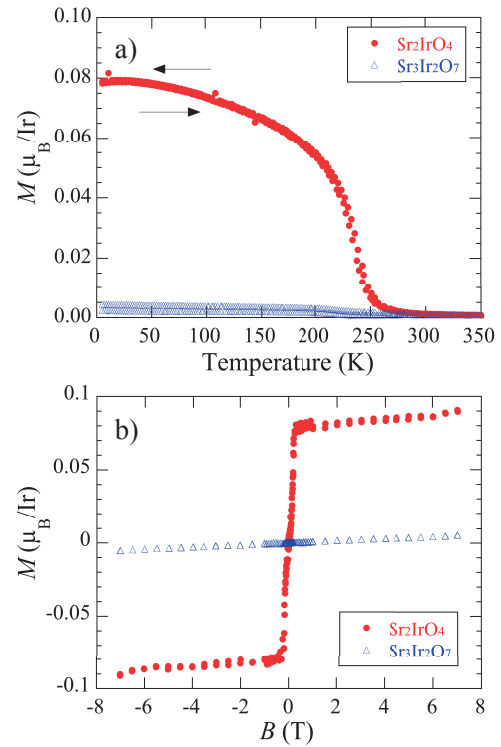}
	\caption{(Color online)   Magnetization ($M$) of \sro\ and \sros\ samples used for $\mu$SR measurements (quoted from Ref.\onlinecite{Fujiyama:12}). (a) $M$  vs.~temperature, where measurements were made after zero-field cooling to 5 K with a sequence of 1) warming up to 350 K under a field $H=1$ T parallel to the $a$-axis and 2) then cooling down to 5 K. (b) $M$ vs.~external field  measured at 5 K ($H\parallel a$). } \label{bchi}
\end{figure}

Polycrystalline samples (mosaic of small single crystals) of \sro\ and \sros\ were prepared using a flux method (with SrCl$_2$ as the flux), while the orthorhombic phase of \sroi\ was obtained by synthesis under a high pressure of 5 GPa.\cite{Longo:71,Ohashi:10}    Structural analysis was performed using the powder x-ray diffraction technique to confirm that these samples mostly comprised a single phase with negligible content of impurities.  The present samples of \sro\ and \sros\ were prepared in the same batch as those measured by Fujiyama {\sl et al}.~using resonant x-ray diffraction.\cite{Fujiyama:12}  The bulk properties of these samples were investigated with different methods, including measurements of the magnetization, electrical resistivity, Hall coefficient, and specific heat, to ensure the sample quality. Fig.~\ref{bchi} shows the results of the magnetization measurements. 

Fig.~\ref{bchi}a clearly shows that a weak ferromagnetic behavior developed below $\sim$250 K in \sro, which is consistent with earlier reports.\cite{Crawford:94,Shimura:95,Cao:98}  This is now understood to come from the canted Ir moments that exhibit a metamagnetic transition at $\sim$0.2 T.\cite{Kim:09}  Meanwhile, \sros\ exhibited much lower magnetization, although the temperature dependence was similar to that in \sro.  \sros\ crystals are known to occasionally contain IrO$_2$ monolayers (corresponding to \sro) as an impurity phase.  Thus, the low magnetization in \sros\ may be attributed to such an impurity phase; its fractional yield was empirically estimated to be less than 0.1\% in single crystals.\cite{Fujiyama:15}  

In contrast, the magnetic susceptibility in the orthorhombic \sroi\ was found to be mostly independent of temperature ($\chi\simeq3.3\times10^{-4}$ emu/mol at 1 T, except for divergent behavior at low temperatures that may be attributed to paramagnetism associated with weak localization below $\sim$45 K). The behavior is understood to be Pauli paramagnetism.\cite{Ohashi:10}
Because the electronic specific coefficient is close to that of conventional metals ($\gamma=$ 2--5 mJ/molK$^2$), the compound has a Wilson ratio $R_{\rm w}=9.8$ that is much greater than the limit of strong correlation ($R_{\rm w}=2$).\cite{Ohashi:10}  Such behavior is consistent with earlier reports\cite{Zhao:08} and was attributed to a large Van Vleck term or strong magnetic correlation. 

\subsection{$\mu$SR Experiment and Data Analysis}
Conventional $\mu$SR measurements were performed using the
Lampf spectrometer installed on the M15/M20 beamlines at TRIUMF, Canada. 
 During measurements under a zero external field (ZF), the residual magnetic field at the sample position was reduced to below $10^{-6}$~T, while the initial muon spin direction was parallel to the muon beam direction
[$\vec{P}_\mu(0)\parallel \hat{z}$].  For longitudinal field (LF)
measurements, a magnetic field was applied parallel to $\vec{P}_\mu(0)$. Time-dependent muon polarization [$G_z(t)=\hat{z}\cdot \vec{P}_\mu(t)$] was monitored by measuring the decay-positron asymmetry along the $\hat{z}$-axis:
\begin{equation}
A(t)\simeq A_0G_z(t)=\frac{N_+(t)-\alpha N_-(t)}{N_+(t)+\alpha N_-(t)}
\end{equation}
where $A_0$ is the average asymmetry. 
$$N_\pm(t)= N_\pm(0)e^{-t/\tau_\mu}[1\pm A_\pm G_z(t)]$$ 
is the positron event rate for the detector placed in the forward ($+$) or backward ($-$) position relative to the sample, $\tau_\mu$ is the muon decay lifetime ($=2.198\times10^{-6}$ s), $A_\pm$ is the decay positron asymmetry for the detector in question ($A_\pm\simeq A_0$), and $\alpha$ is the instrumental asymmetry [$\alpha=N_+(0)/N_-(0)\simeq1$ under normal conditions]. 

\begin{figure}[tb]
	\centering
	\includegraphics[width=0.95\linewidth,clip]{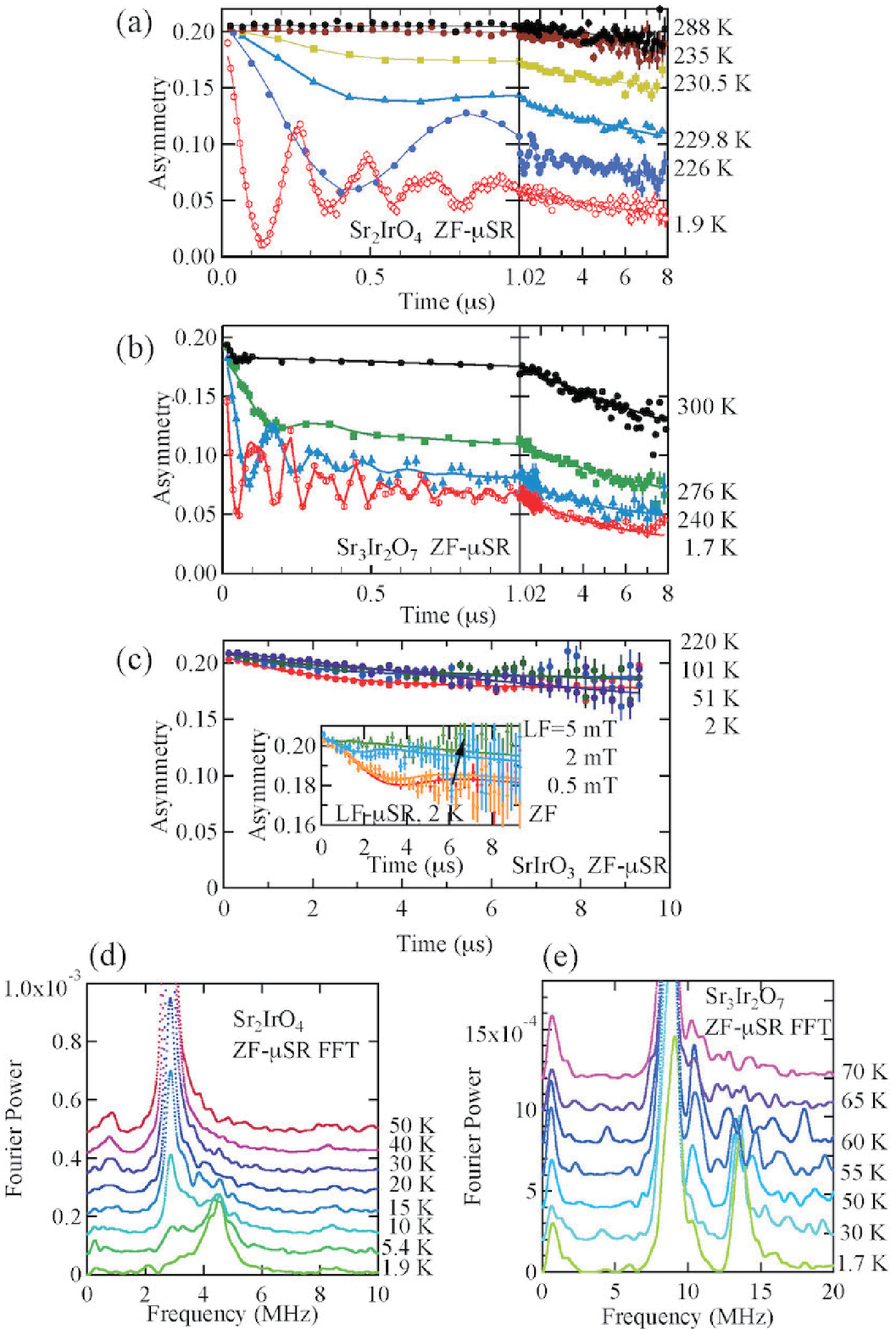}
	\caption{(Color online) Typical \msr\ time spectra under zero external field (ZF) in  (a) \sro, (b) \sros\, and (c) \sroi, with inset showing some spectra under longitudinal field (LF). Modulation of sinusoidal pattern due to overlap of signals with different frequencies [$\sim$4.5 MHz and $\sim$8 MHz at 1.9 K in (a), $\sim$9 MHz and $\sim$13 MHz at 1.7 K in (b)] can be clearly observed in the earlier time range of the spectra up to 1 $\mu$s. Fast Fourier transform of ZF-\msr\ time spectra is shown for (d) \sro\ and (e) \sros. (f) LF-\msr\ spectra in \sros.} \label{tspec}
\end{figure}

Fig.~\ref{tspec} shows several examples of time-dependent ZF/LF-\msr\ asymmetry spectra and their fast Fourier transform (FFT)  in \srom; those for $n=1$ and 2 exhibit  a clear sinusoidal oscillation below $T_{\rm C}$.  Such a signal indicates the onset of a relatively homogeneous internal field at muon sites in accordance with the long-range magnetic order.  The absence of enhanced muon depolarization just above $T_{\rm C}$ (usually expected as a consequence of the critical slowing down of the spin fluctuation) and the volumetric expansion of fraction for the magnetically ordered region below $T_{\rm C}$ (as inferred from the increase in oscillation amplitude with decreasing temperature) suggest that the transition is driven not by the magnetic interaction but by the Mott transition.  In contrast, the spectra in \sroi\ ($n=\infty$) did not exhibit spontaneous oscillation over the entire observed range of the temperature, which demonstrates that no long-range magnetic order is present above 2 K.

More interestingly, closer investigation of the spectra for the former two cases revealed a secondary frequency component at lower temperatures, which can be clearly observed in the Fourier transform shown in Fig.~\ref{tspec}(d) and (e).  The signal amplitude of the secondary component increased instead of the first one, which suggests the gradual development of the former with decreasing temperature. Considering this, we analyzed the ZF-\msr\ time spectra by curve-fitting using the following model function which is valid for the ordered phase of polycrystalline samples:
\begin{eqnarray}
A(t)&\simeq&\sum_{i=1}^m A_i\left[\frac{1}{3}e^{-\lambda_{\rm L}t}+\frac{2}{3}G_z(t,\Delta) e^{-\lambda_it}\cos(\omega_it+\phi)\right]\nonumber\\
& &\hspace{9em}+A_{\rm para}G_z(t,\Delta)  \label{fit}
\end{eqnarray}
where $G_z(t,\Delta)$ is the Kubo-Toyabe relaxation function\cite{Hayano:79} to describe the slow Gaussian depolarization due to random local fields exerted from {\sl nuclear} magnetic moments (with $\Delta\le10^{-1}~\mu {\rm s}^{-1}$ being the linewidth in the quasistatic limit), $A_i$ is the partial asymmetry for the magnetic phase (with $\sum_iA_i=A_{\rm mag}$), $A_{\rm para}$ is the asymmetry for the paramagnetic phase, $\lambda_{\rm L}$ is the longitudinal relaxation rate, $\lambda_i$ is the transverse relaxation rate, $\omega_i$ is the muon spin precession frequency ($=2\pi f_i=\gamma_\mu B_i$ with $\gamma_\mu=2\pi\times135.53$ MHz/T and $B_i$ being the local field at the muon site), and $\phi$ is the initial phase of precession.  Note that the first term in Eq.~(\ref{fit}) represents the situation where the Kubo-Toyabe depolarization would be quenched [{\it i.e.,} $G_z(t,\Delta)\simeq1$] for  one third of the implanted muons exposed to a longitudinal internal field in the magnetic phase. 
We found that four components ($m=4$) were sufficient to describe the spectra observed in \sro. 
Similarly, the spectra in \sros\ required four components with one showing no oscillation, while one of the rest showed further splitting  ($m=5$) when $T\le T_{\rm m}$.

\section{Result}
\subsection{\sro}
\begin{figure}[t]
	\centering
	\includegraphics[width=0.9\linewidth,clip]{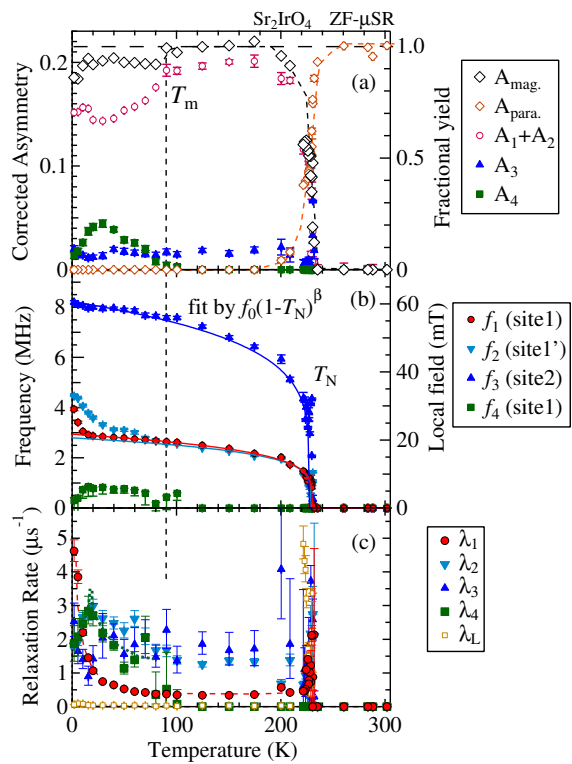}
	\caption{(Color online) Temperature dependence of (a) partial asymmetry $A_i$, $A_{\rm para}$, (b) muon precession frequency $f_i$, and (c) relaxation rate $\lambda_i$, $\lambda_L$ in \sro. Solid curves are results of curve-fitting by $f_0(1-T/T_{\rm N})^\beta$, and dashed curves are visual guides. For the correspondence between $f_i$ and muon sites, see Fig.~\ref{musite1}.} \label{param}
\end{figure}

Fig.~\ref{param} shows the partial asymmetry ($A_i$, $A_{\rm para}$), muon precession frequency ($f_i$), and relaxation rate  ($\lambda_i$) as determined from the curve-fitting of ZF-\msr\ spectra in \sro\ as a function of temperature.  As discussed below, the splitting of signals into multiple frequencies can be attributed to the statistical distribution of implanted muons into magnetically inequivalent interstitial sites in the unit cell (situated at the local minima of electrostatic potential), where muons are exposed to different internal fields. The muon-electron hyperfine interaction is predominantly determined by the short-range magnetic dipolar interaction; thus, it strongly depends on the local configuration of the nearest neighboring magnetic ions.    
The onset temperature ($T_{\rm N}$) for the two precessing components $f_1$ ($=f_2$ for $T>T_{\rm m}$) and $f_3$ agreed well with the $T_{\rm C}\simeq240$ K suggested by bulk magnetization.\cite{Crawford:94,Shimura:95,Cao:98}  $T_{\rm N}$ also perfectly matched with the values inferred from resonant x-ray diffraction (RXD).\cite{Kim:09}  Curve-fitting of the frequency by the power law $f_i(T)=f_{(0)i}(1-T/T_{\rm N})^\beta$ yielded the critical index $\beta\simeq0.2$ (see Table \ref{tab214} for more detail) which is situated somewhere between the three-dimensional (3D) Heisenberg  ($\beta=0.365$) and two-dimensional (2D) Ising ($\beta=0.125$) spin systems.
Note that $\beta$ is distinctly smaller than that observed by RXD\cite{Kim:09} and neutron\cite{Ye:13} [$\beta\simeq0.35$ ($\equiv2\beta$ in Ref.\onlinecite{Ye:13})], which may be attributed to the volumetric expansion of magnetic domains with decreasing temperature as suggested by \msr\ [see $A_i$ in Fig.~\ref{param}]. The intensity of x-ray/neutron diffraction should be  proportional to the product of $f_i$ ($\propto$ the true order parameter) and $A_i$ ($\propto$ volumetric fraction of magnetic domains).  Thus, precautions must be taken when interpreting diffraction data in terms of critical indices for the magnetic order parameter.

Nonetheless, the most salient feature in Fig.~\ref{param} is the gradual departure of $f_{1,2}$ from the curve predicted by the power law and further splitting into two frequencies ($f_1 < f_2$)  below $T_{\rm m}\simeq 90$ K; it shows a much steeper increase as the temperature decreases below 20--30 K.  A corresponding behavior is suggested for $f_3$, for which the relaxation rate increased below $T_{\rm m}$. An additional component ($i=4$) showing strongly damped precession  ($\lambda_4\ge f_4$) also developed. The signal was accompanied by a loss of the initial asymmetry (seen from the reduction of $A_{\rm mag}$ by $\sim$10\%, suggesting fast depolarization over a time range of $10^{-9}$ s).  A similar $\mu$SR result was reported in a previous paper with some different details with regard to the behavior of fitting parameters.\cite{Franke:11} 

Surprisingly similar behavior has previously been observed for the muon precession frequency in another Ir perovskite \bro, except for the minor difference that the signal corresponding to $f_3$ is missing.\cite{Okabe:11} The difference may be due to the difficulty in identifying the precession signal of relatively small amplitude ($A_3\simeq$7--12\% in \sro), although it may also be due to the change in the population of muon sites induced by the modulation of the crystal structure upon the substitution of Sr by Ba.

\subsection{\sros}
\begin{figure}[t]
	\centering
	\includegraphics[width=0.9\linewidth,clip]{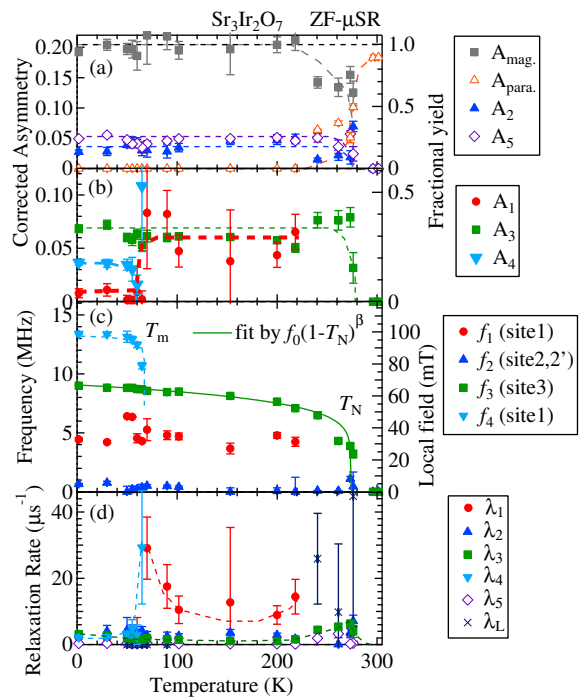}
	\caption{(Color online) Temperature dependence of (a) (b) partial asymmetry, (c) muon precession frequency and (d) relaxation rate in \sros. Solid curve is fitting line, and dashed curves are visual guides. For the correspondence between $f_i$ and muon sites, see Fig.~\ref{musite2}.} \label{param2}
\end{figure}

The ZF-\msr\ spectra in \sros\ were described by four components ($i=1$--3, 5) for $T_{\rm m}<T<T_{\rm N}$, while another component ($i=4$) appeared below $T_{\rm m}$.  Such a two-step behavior is qualitatively parallel to that in \sro.  Fig.~\ref{param2} shows the temperature dependence of $A_i$, $f_i$, and $\lambda_i$, where $f_3$ seems to be the primary component reflecting the magnetic order parameter for $T_{\rm m}<T<T_{\rm N}$, while the behavior of the other components ($f_1\simeq4$ MHz, $f_2\simeq0.6$ MHz, and $f_5\simeq0$ MHz) is relatively ambiguous.  The transition temperature $T_{\rm N}=273.7(4)$ K determined by the onset of $f_3$ showed close agreement with the bulk properties, including the specific heat.\cite{Cao:02,Nagai:07}  Curve-fitting of $f_3$ using $f_3(T)=f_{(0)3}(1-T/T_{\rm N})^\beta$ yielded $\beta\simeq0.143(3)$, which is closer to the value for the 2D-Ising model ($\beta\sim0.125$) than the case of \sro. This suggests a stronger 2D characteristic of the magnetic order in \sros. 

Although the value of $f_{(0)3}$ [$=9.07(3)$ MHz] happened to be close to $f_{(0)3}$ in \sro\ [$=8.22(4)$ MHz], the $T_{\rm N}$ values determined by these signals clearly differed between the respective compounds. (This in turn indicates that the $f_3$ component in the \sro\ sample cannot be attributed to the \sros\ impurity phase.) 

From the appearance of $f_4$ below $T_{\rm m}\simeq70 $ K, \sros\ was inferred to undergo a secondary magnetic transition for $T<T_{\rm m}$.  The transition was accompanied by a reduction in $A_1$,  which suggests that the signal comes from muons formerly associated with $f_1$ for $T>T_{\rm m}$. Furthermore, both $\lambda_1$ and $\lambda_4$ exhibited a steep increase around $T_{\rm m}$, which is regarded as the critical divergence of the relaxation rate often observed for the conventional magnetic order. Figs.~\ref{tspec}(d) and (e) show that the relative amplitude of $f_4$ ($\propto A_4$) was mostly independent of the temperature in \sros\ below $T_{\rm m}$. This is in contrast to the case in \sro, where it increased as the temperature decreased  below $\sim$30 K. These results suggest that these two phases (or ``states") coexist in \sros\ below $T_{\rm m}$. Thus, while the magnetism of \sros\ shares the feature of a secondary phase below  $T_{\rm m}$ with \sro, the temperature dependence of its order parameter demonstrates one of the differences between these two compounds. 

The above results are substantially different from that reported earlier in the literature.\cite{Franke:11}  Franke {\sl et al.} reported emergence of a single frequency component over a temperature range of $25 \le T \le160$ K, where the magnitude of the frequency (1.8--2.8 MHz) and its temperature dependence were rather close to that of $f_1$ in \sro. In contrast, we found no corresponding signal in the present result for \sros. They also refered to a signal component that exhibits extremely rapid relaxation ($\sim$40 MHz) over the entire temperature range that they studied ($\le260$ K), which is missing from our result.  These observations suggest that the specimen used by Franke {\sl et al.}\cite{Franke:11} may have contained a significant fraction of ingrowth \sro\ phase upon preparation.

\subsection{\sroi}
Unlike the preceding two cases, \sroi\ remained non-magnetic over the entire temperature range studied (2 K$<T<250$ K), which was inferred from the absence of a spontaneous muon precession signal in ZF-\msr\ spectra. More specifically, the spectra shown in Fig.~\ref{tspec}(c) are the sum of two components showing Gaussian depolarization due to nuclear random local fields. 
While the depolarization was mostly quenched by applying a longitudinal magnetic field of $10^1$ mT,  a slowly depolarizing component remained [see the inset of Fig.~\ref{tspec}(c) showing the data at 2 K].  This suggests the presence of residual fast spin dynamics of Ir 5$d$ electrons behind the random local fields from nuclear dipolar moments.
The spectra were analyzed by curve-fitting using the form 
\begin{equation}
A(t)=\sum_{i=1}^2 A_i G_z(t,\Delta_i)e^{-\lambda_it} \\ \label{fit2}
\end{equation}
which becomes
\begin{equation}
\simeq \sum_{i=1}^2 A_i \left[\frac{1}{3}+\frac{2}{3}(1-\Delta_i^2t^2) e^{-\Delta_i^2t^2/2}\right]e^{-\lambda_it}\nonumber
\end{equation}
under a zero field, 
where $\Delta_i$ is the Gaussian relaxation rate determined by the second moment of nuclear dipolar fields and $\lambda_i$ is the relaxation rate due to the fluctuation of paramagnetic moments.
We obtained $\Delta_1=0.026(2)~\mu{\rm s}^{-1}$ and $\Delta_2\simeq0.46(2)~\mu{\rm s}^{-1}$ from the fitted curves, including those under LF-$\mu$SR measured at 2 K. 

Fig.~\ref{tspec}(c) shows that the primary component of the time spectra ($A_1\ge0.18$) did not appreciably change with time nor under different LF. This makes it difficult to reliably determine $\Delta_1$ and $\lambda_1$ as independent parameters. A similar difficulty arose for the secondary component from the small signal amplitude ($A_2\le0.02$). 
Consequently, $\lambda_2$ alone was treated as a parameter in the curve-fitting to determine the temperature dependence of the spectra, while $\Delta_{1,2}$ was fixed to the values determined at 2 K  (except for $T>200$ K  where $\Delta_2$ was set free) and $\lambda_1$ was fixed to zero. In such a situation, $\lambda_2$ extracted from curve-fitting would be equivalent to $\lambda_1\simeq(A_2/A_1)\cdot\lambda_2$ under the reverse conditions.

\begin{figure}[t]
	\centering
	\includegraphics[width=0.7\linewidth,clip]{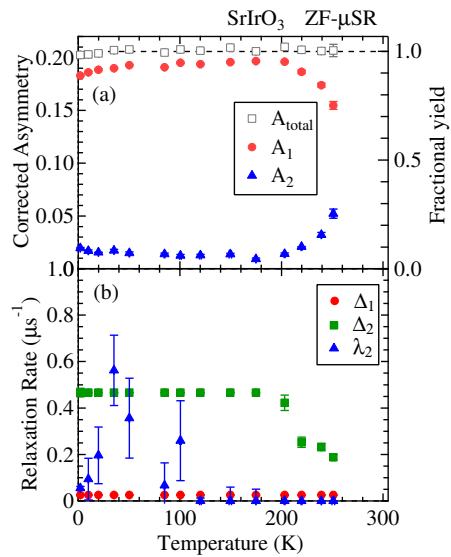}
	\caption{(Color online) Temperature dependence of (a) partial asymmetry $A_i$ and (b) relaxation rate $\Delta_i$, $\lambda_2$ in \sroi.} \label{param3}
\end{figure}

Fig.~\ref{param3} summarizes the analysis results.  $\lambda_2$ exhibited a weak enhancement around 30--50 K, which may suggest a slowing down of paramagnetic spin fluctuation. This is consistent with the bulk magnetic properties, which are mostly due to paramagnetism. The electronic structure of \sroi\  is reportedly on the verge of the metal-insulator transition, so the conducting carriers would have a large effective mass because of the strong electronic correlation.\cite{Moon:08}  At this stage, there were a number of bulk property measurements with anomalies around 30--50 K: the resistivity, Seebeck, and Hall coefficients exhibited a kink at $\sim$45 K, and the Nernst effect showed a peak at $\sim$35 K. These anomalies seem to coincide with the enhancement of $\lambda_2$ observed over a similar temperature range.

\section{Discussion}

In the magnetically ordered states of \sro\ and \sros, the Ir$^{4+}$ isospin state is virtually equivalent to the superposition of a magnetic dipole and octupole for $f$ electron systems, where the octupole makes a relatively small contribution to reducing the total magnetic induction.  Thus, the local field at the muon site can reasonably be assumed to be predominantly determined by the point-like magnetic dipoles situated at Ir sites. 
The internal field $B_{\rm loc}$ at the muon site is given by the vector sum of the dipolar fields exerted from  local Ir$^{4+}$ isospin moments: 
 \begin{equation}
B_{\rm loc} =|\sum_j{\bf \hat{A}}_j{\bm \mu}_j|
\end{equation}
 where ${\bf \hat{A}}_j$ is the dipole tensor and is expressed by 
\begin{equation}
{\bf \hat{A}}_j=A^{\alpha\beta}_j=\frac1{r_j^3}\left(\frac{3\alpha_j\beta_j}{r_j^2}-\delta_{\alpha\beta}\right)\quad(\alpha, \beta=x,y,z).
\label{dip}
\end{equation}
The summation runs through the $j$-th Ir$^{4+}$ moment ${\bm \mu}_j$ located at ${\bm r}_j=(x_j,y_j,z_j)$ from a given muon site.  
For a given magnetic structure and  Ir$^{4+}$ moment size,  $B_{\rm loc}$  can be calculated for comparison with the observed $$B_i=2\pi f_i/\gamma_\mu.$$  

The local electronic state associated with a muon is virtually identical with that of hydrogen as an interstitial atom because the difference in the reduced mass of the electron bound to the muon (which determines the Bohr atomic radius) is only $\sim$0.5\% compared with that of hydrogen.  Moreover, in transition metal (TM) oxides, the energy level of the electronic state associated with interstitial hydrogen $E^{+/-}$ has been empirically established to be predominantly determined by a common offset of nearly $-3$ eV measured from the vacuum level regardless of the host compounds;\cite{Kilic:02,Cox:06} the muon is predicted to stay diamagnetic (accompanying no unpaired spin density) as long as $E^{+/-}>E_{\rm c}$ (where $E_{\rm c}$ is the energy at the bottom of the conduction band). This is perfectly in line with the common observation that the muon in TM oxides satisfying the condition $E^{+/-}>E_{\rm c}$ tends to form an O-$\mu$ covalent bond (a muonic analogue of OH bond), where the local  O-$\mu$ binding energy is the primary factor that determines $E^{+/-}$.  Because \srom\ has a relatively narrow band gap (i.e., $E_{\rm c}$ is far below the vacuum level, so $E^{+/-}>E_c$), we can safely presume that the muon forms an O-$\mu$ bond in these Ir compounds.
Accordingly, we searched for muon sites bound to oxygen (with a typical O-$\mu$ distance of $\sim$0.1 nm) in the respective compounds that are situated near the (local) minima of the electrostatic potential that are induced by local ions.

The electrostatic potential was calculated using a custom-made computer program assuming a point charge for each atom,\cite{Kojima:04} where the valence state and lattice structure were the input parameters.  In addition, the Vienna Ab-initio Simulation Package\cite{VASP} (VASP, codes for the ab-initio quantum mechanical molecular dynamics calculation) was used to examine the validity of the potential minima derived from the calculation based on the point-charge approximation.  These calculations showed perfect agreement for the case of \sro, while a slight difference was suggested for \sros\ and \sroi\ (see below).

In the paramagnetic state, the muon-Ir$^{4+}$ hyperfine parameter (which is predominantly determined by the magnetic dipolar interaction) is estimated by calculating the second moment
\begin{equation}
A_\mu^2=\sum_{j,\alpha,\beta}\left[(A^{\alpha\beta}_j)\mu^\beta_j\right]^2~~(\alpha=x,y,\:\beta=x,y,z), 
\label{dlt}
\end{equation}
where the sum runs over the $\alpha$ and $\beta$ components of the terms exerting a magnetic field perpendicular to initial muon polarization from the $j$-th nuclear magnetic moment.
The nuclear dipolar width ($\Delta^2/\gamma_\mu^2$) can be estimated with the above equation by substituting ${\bm \mu}_j$ with those of the nuclear magnetic moments.

In the following, the assignment of muon sites is discussed through a comparison of $B_i$ and the calculated values of $B_{\rm loc}$ in \sro\ and \sros. In contrast,  it is solely based on $\Delta$ in \sroi\ because the compound remained paramagnetic for the entire temperature range considered in the present study.
\begin{table*}[t]
	\caption{Muon spin precession frequency (extrapolated to $T=0$ by curve-fitting using the form $f_{(0)i}(1-T/T_{\rm N})^\beta$ for $T_{\rm m}\le T\le T_{\rm N}$, where $T_{\rm N}$ is the N\'eel temperature) and corresponding internal field ($B_i$) in the AF phase of \sro. $T_{\rm N}$ and the critical index $\beta$ are determined by the fitted curves. The Ir moment size is estimated using the equation $\mu_{\rm Ir}=B_i/\overline{B}_{\rm loc}$, where $\overline{B}_{\rm loc}$ is the field calculated for the Ir moment size of 1$\mu_B$. Muon sites were assigned by searching for the positions of local potential minima around oxygen (with a distance of 0.1 nm),  where the corresponding $\mu_{\rm Ir}$ falls within a variation of 10\% from the known value of 0.4$\mu_B$.  (The position is given in polar coordinates with the $\hat{z}$ axis parallel to the crystalline $c$ axis.)  O1: apical oxygen (0.5, 0.25, 0.2047); O2: in-plane oxygen (0.2001, 0.4501, 0.125); O2': in-plane oxygen (0.7001, 0.5499, 0.125) \label{tab214}}
\begin{tabular}{ccccccccc} \hline\hline
  & & & & & & & &  \\[-8pt]
Site\# &	     $i$   & $f_{(0)i}$ (MHz) & $B_i$ (mT) & $T_{\rm N}$ (K) & $\beta$ &  $\mu^+$ position ($\theta, \phi$) & $\overline{B}_{\rm loc}$ (mT/$\mu_B$) & $\mu_{\rm Ir}$  \\ \hline 
 & & & & & & & &  \\[-8pt]
Site1   & 1 & 2.941(3) &   21.70(2) & 230.1(1) & 0.211(1) & O1+(60$^\circ$, 45$^\circ$) 	&  51.6 & 0.4205(4)$\mu_B$  \\
Site1' &  2 & 2.79(4) & 20.6(3) & 229.0(2) & 0.192(5) & 	O1+(60$^\circ$, 135$^\circ$)	& 53.0 & 0.389(6)$\mu_B$ \\
 Site2 & 3 & 8.22(4) & 60.6(3)	& 227.4(3) & 0.190(7) & O2+($\sim$16$^\circ$, 45$^\circ$) & 142.4  & 0.426(2)$\mu_B$\\
 Site2' &	   	-- & -- & --	& -- & --& O2'+($\sim$16$^\circ$,135$^\circ$)		& 162.1 & -- \\
  & & & & & & & &  \\[-8pt]
	  \hline\hline
	\end{tabular}
\end{table*}

\subsection{\sro}

\begin{figure}[t]
	\centering
	\includegraphics[width=0.9\linewidth,clip]{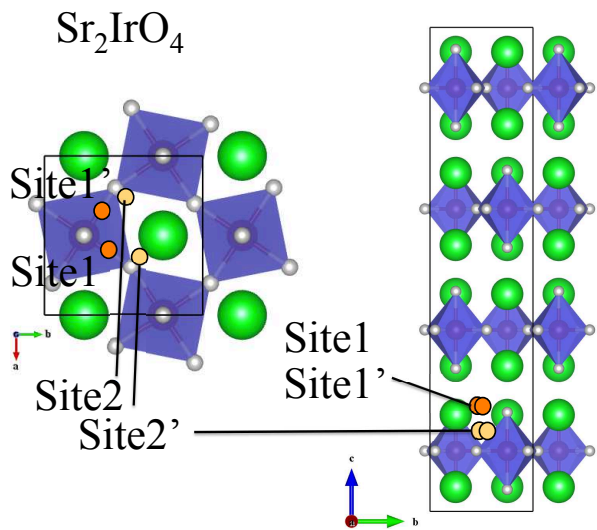}
	\caption{(Color online) Presumed muon sites in \sro. Site1 and Site1' are close to apical oxygen (O1), and Site2 and Site2' are near the IrO$_2$ plane (O2). } \label{musite1}
\end{figure}

We adopted a canted AF structure (with ${\bm \mu}_j$ tilted by $11^\circ$ from the $a$ axis) and the Ir moment size determined from RXD and neutron diffraction measurements ($|{\bm \mu}_j|\simeq0.36\mu_B$) in \sro. We further assumed that the relevant magnetic structure is realized over the temperature region $T_{\rm m}< T<T_{\rm N}$.  Then, considering that there are two crystallographically inequivalent oxygen sites, i.e., apical oxygen (O1) and in-plane oxygen (O2), the comparison between $f_i$ and $f_\mu\equiv\gamma_\mu B_{\rm loc}$ yields a consistent set of muon sites. This is shown in Fig.~\ref{musite1}, where $f_1$ and $f_2$ correspond to muons bound to O1 (Site1 and Site1'), whereas $f_3$ is associated with those bound to O2 (Site2) (see Table \ref{tab214} for more details).  The magnitude of $A_i$ represents the initial distribution of implanted muons among the corresponding sites. The Ir$^{4+}$ moment size is commonly estimated to be $\sim$0.4$\mu_B$ from the magnitude of frequencies for all muon sites, which closely agrees with the value inferred from RXD data for $T_{\rm m}<T<T_{\rm N}$.

For the in-plane oxygen (O2), our calculation of the electrostatic potential predicted another set of sites with lower potential energy than Site2. The corresponding hyperfine parameters calculated from the dipole tensor were 256.4/248.1 mT/$\mu_B$. Thus, a \msr\ frequency of 13.9/13.5 MHz is predicted for the Ir moment size of $\sim$0.4$\mu_B$.  We presumed that the corresponding \msr\ signal may not have been resolved in the present \msr\ time spectra because of the small amplitude and/or fast depolarization based on the achieved statistical precision. The small reduction of the total asymmetry ($A_{\rm mag}$) observed below $T_{\rm m}$ (see Fig.~\ref{param}) also suggests that an unresolved (missing) signal exists. 

These site assignments with the presumed Ir moment size do not necessarily correlate with the distinct features observed below $T_{\rm m}$ (see Fig.~\ref{param}): (i) $f_1$ splits into two lines, and (ii) they exhibit a divergent increase below $\sim$40 K.  It is unlikely that feature (i) can be explained by further canting of the Ir isospin moments (i.e., from $\sim$11$^\circ$ to $\sim$13$^\circ$ within the $ab$ plane inferred from x-ray and neutron diffraction,\cite{Kim:09,Ye:13}) because the expected change in $f_{1,2}$ induced by such a canting ($\sim$0.088 MHz) would be too small to explain the observed change. We also investigated a variety of possibilities for magnetic structures, including the spin-flop from the $ab$-in-plane to the $c$-parallel configuration (as has been suggested to occur upon Mn substitution for Ir by 10\%).\cite{Calder:12} However, we found that the calculated $f_\mu$ was far below the observed change of $f_{1,2}$ up to 4--5 MHz for any arbitrary spin configuration.

The change of $f_{1,2}$ below $\sim$40 K also cannot be attributed to the increase of the Ir isospin moment size (without a change in the magnetic structure), because this would lead to an enhancement of $f_{3}$, which is also proportional to the Ir moment size ($\propto$ order parameter).  Fig.~\ref{param} clearly shows that $f_3$ does not exhibit divergent behavior that is in marked contrast to $f_{1,2}$. 

These observations have led us to postulate that the magnetism below $T_{\rm m}$ is associated with the development of  spin polarization at the relevant oxygen sites.  We discuss this possibility in the next section  in detail.

\subsection{\sros}

\begin{figure}[t]
	\centering
	\includegraphics[width=1.0\linewidth,clip]{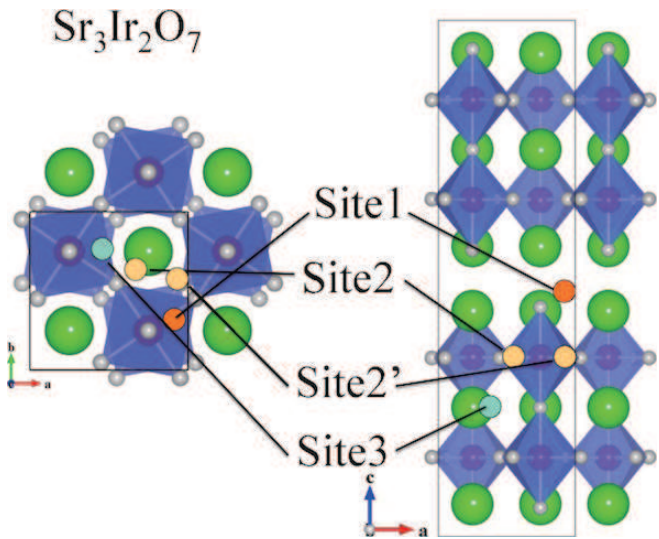}
	\caption{(Color online) Presumed muon sites in \sros. Site1 is close to apical oxygen (O1), Site2 and 2' are near IrO$_2$ planes (O2,O2'), and Site3 is near the shared apical oxygen (O3).} \label{musite2}
\end{figure}

There are three crystallographically inequivalent oxygen sites in \sros:  the outer apical oxygen (O1), IrO$_2$ in-plane oxygen (O2), and inner apical oxygen shared by two IrO$_6$ octahedra (O3). Assuming the occurrence of an O-$\mu$ bond common to the case in \sro, we investigated potential minima for the $\mu^+$ around each oxygen that also meets the criterion that the Ir moment sizes estimated from the magnitude of $B_{\rm loc}$ for different sites agrees with each other (given that the Ir moment size is uniform).  Fig.~\ref{musite2} shows the results, where $f_1$ corresponds to O1 (Site1, near the outer apical O), $f_2$ corresponds to O2 (Site2, close to in-plane O), and $f_3$ corresponds to muons bound to O3 (Site3, near the shared apical O; refer to Table \ref{tab327} for more detail).  We also note that $f_2$ tended to converge to two different frequencies, although it was difficult to extract these two values simultaneously from curve-fitting. We interpreted this as a trace of a signal from another site near O2 (Site2') that was barely resolved. Interestingly, the VASP calculation suggests that Site2' is 0.13 nm away from O2 (whereas the point charge approximation suggests 0.1 nm), which would reduce $B_{\rm loc}$ to 14.3 mT. However, such a small difference would be hard to identify in the present data. 

In addition, we tested three possible AF structures, i.e., those collinear with one of  three crystal axes ($a$, $b$, or $c$), and found that the one collinear with the $c$-axis met the criterion.   The structure showed perfect agreement with that inferred from RXD measurements,\cite{JWKim:12,Fujiyama:12} which supports our assumption that the muon sites were correctly identified. Consequently, the Ir moment size was determined to be $\sim$0.31(2)$\mu_B$ from the average $\mu_{\rm Ir}$ for Site1-3. 

At this stage, the origin of the secondary transition at $T_{\rm m} = 70$ K inferred from the appearance of $f_4$ is unclear. While the amplitude of $f_4$ ($A_4$) developed in place of that for $f_1$ ($A_1$, the signal from muons bound to the apical oxygen), it showed no correlation with other frequency signals that remained unchanged as the temperature passed through $T_{\rm m}$. As is the case for \sro\ below $T_{\rm m}$, neither the change in the Ir moment orientation nor in its moment size can explain such a behavior.  Attempts to attribute the observed change to a variety of modulated Ir moment orientation/size always lead to a change in $B_{\rm loc}$ for all muon sites, which is inconsistent with the current experimental observation. 

The development of anomalous	 internal field is commonly observed with muons at the near-apical oxygen site in both \sro\ and \sros.  As suggested previously, we now speculate that this is in accordance with the development of ordered magnetic moments at the relevant oxygen site, which is common to these two compounds (see below). 

\begin{table*}[t]
\caption{Muon spin precession frequency ($f_i$) in the AF phase of \sros. While $f_3$ can be extrapolated to $T=0$ by curve-fitting using the form $f_{(0)3}\cdot(1-T/T_{\rm N})^\beta$ for $T_{\rm m}\le T\le T_{\rm N}$), other components only allow a simple linear extrapolation with considerable uncertainty because of the relatively large scattering of data (see Fig.~\ref{param2}). Regarding the double value of $f_2$, refer to the main text. The Ir moment size is estimated with the equation $\mu_{\rm Ir}=B_{\rm loc}/\overline{B}_{\rm loc}$, where $\overline{B}_{\rm loc}$ is the field calculated for the Ir moment size of 1$\mu_B$. Muon sites were assigned by searching for the position of local potential minima around oxygen (with a distance of 0.1 nm),  where the corresponding $\mu_{\rm Ir}$ are reasonably close to each other.  (The position is given in polar coordinates with the $\hat{z}$ axis parallel to the  crystalline $c$ axis.)  O1:  outer apical oxygen (0.75, 0.25, 0.1943); O2:  in-plane oxygen (0.5487, 0.5485, 0.0962); O2':  in-plane oxygen (0.0487, 0.4515, 0.3462);  O3: shared apical oxygen (0.5, 0.25, 0) \label{tab327}}
\begin{tabular}{ccccccccc} \hline\hline
  & & & & & & & &  \\[-8pt]
Site\# & $i$ & $f_{(0)i}$ (MHz) & $B_i$ (mT) & $T_{\rm N}$ (K) & $\beta$ & $\mu^+$ position ($\theta, \phi$) & $\overline{B}_{\rm loc}$ (mT/$\mu_B$) &  $\mu_{\rm Ir}$ \\ \hline
 & & & & & & & &  \\[-8pt]
Site1  & 1 & 4.5(5)  & 33.2(37) & -- & -- & O1+(60$^\circ$,45$^\circ$)	 & 92.0 & 0.36(4)$\mu_B$ \\
Site2 & 2 & 0.30(8) & 2.2(6) & -- & -- & O2+(90$^\circ$,45$^\circ$) & 8.06 & 0.28(7)$\mu_B$\\
Site2' & 2 & 0.70(5) & 5.2(4) & -- & -- & O2'+(90$^\circ$,135$^\circ$) & 17.9 &   0.29(2)$\mu_B$\\
Site3 & $3$ & 9.07(3) & 66.9(2) & 273.7(4) & 0.143(3) & O3+(90$^\circ$,45$^\circ$) &  215.5 & 0.310(1)$\mu_B$\\ 
 & & & & & & & &  \\[-8pt]
\hline\hline
	\end{tabular}
\end{table*}

\subsection{\sroi}

The number of crystallographically inequivalent oxygen is reduced to two in the orthorhombic \sroi, which is similar to \sro\ except that the apical oxygen (O1) is shared by two IrO$_6$ octahedra.  In this situation,  there are two possible muon sites for the presumed O-$\mu$ bonding, just like for \sro.  The muon sites were searched for by calculating the electrostatic potential using lattice parameter and atomic coordinates for the orthorhombic perovskite  \sroi\ with $Pbnm$ crystal symmetry.\cite{Longo:71,Zhao:08}  We obtained a potential minimum for each oxygen atom (Site1 for O1 and Site2 for O2, see Table \ref{tab113}).  

As noted earlier, the analysis of LF-$\mu$SR data yielded $\Delta_1$=0.026(2) $\mu{\rm s}^{-1}$ and $\Delta_2$=0.46(2) $\mu{\rm s}^{-1}$ for each signal component.  However, the calculated line width turned out to be nearly the same for these sites with $\Delta\simeq0.024$--0.026 $\mu{\rm s}^{-1}$.   We also found that calculation using VASP codes predicted another potential minimum situated at the center of four apical oxygen atoms, for which $\Delta\simeq0.022$ $\mu{\rm s}^{-1}$.
While these values of $\Delta$ closely agrees with $\Delta_1$ (corresponding to the predominant signal), there is a large discrepancy with $\Delta_2$.  Note that  $\Delta$ never exceeded 0.20 $\mu{\rm s}^{-1}$ at any available interstitial site  (regardless of the likelihood of occupancy according to the electrostatic potential). This suggests that the signal corresponding to $\Delta_2$ does not come from \sroi\ but from an unidentified impurity phase.  However, we stress  that the volumetric fraction of this phase was less than $\sim$10\% below $\sim$200 K (see Fig.~\ref{param3}), which indicates that the Pauli paramagnetic behavior can be attributed to the predominant fraction making up the \sroi\ phase.

The muon-Ir hyperfine parameter ($A_\mu$) in \sroi\ as a paramagnetic metal can be estimated by calculating the dipolar sum using Eq.~(\ref{dlt}).  Table \ref{tab113} summarizes the results.  $A_\mu$ was found to take similar values for those two muon sites ($\simeq10^2$ mT/$\mu_B$). Considering that the muon depolarization rate under spin fluctuation can be approximated with the Redfield theory of spin relaxation
\begin{equation}
\lambda\simeq\frac{\gamma_\mu^2A_\mu^2}{\nu}
\end{equation}
and that no appreciable depolarization was observed in this compound within the time window of the observation (i.e., $\lambda<10^{-2}$ $\mu$s$^{-1}$), we may be able to place a lower bound for the Ir spin fluctuation rate:
\begin{equation}
\nu>10^{11}\:\:{\rm s}^{-1}.
\end{equation}
Assuming that \sroi\ can be understood as a normal metal (despite it being rather close to a semimetal with a small number of carriers), the spin fluctuation due to the Pauli paramagnetism can be estimated as
\begin{equation}
\nu\simeq\frac{1}{\hbar D(\varepsilon_F)}=\frac{2\pi^2k_B^2N_A}{3\hbar\gamma_e}\simeq2.9\times10^{12}\:\:{\rm s}^{-1},
\end{equation}
where $D(\varepsilon_F)$ is the density of state at the Fermi level, $N_A$ is the Avogadro number, and $\gamma_e=2.45$ mJ/mol K$^2$  is the Sommerfeld coefficient.\cite{Ohashi:10} This is also consistent with the present experimental result where virtually no appreciable depolarization was observed.

\begin{table}[t]
	\caption{Muon site assignment in \sroi\ made by locating the potential minimum around oxygen (with a distance of 0.1 nm), where the nuclear dipolar line width ($\Delta$) is consistent with experimental observation. (The muon position is given in polar coordinates with the $\hat{z}$ axis parallel to the  crystalline $c$ axis.)   O1: share apical oxygen (0.9940, 0.5073, 0.2500); O2: in-plane oxygen (0.7920, 0.7860, 0.5440)  \label{tab113}}
	\begin{tabular}{cccc} \hline\hline
Site\#  & $\mu^+$ position ($\theta, \phi$) 	& $\Delta_i$ ($\mu$s$^{-1}$)	& $A_\mu$ (mT/$\mu_B$) \\ \hline 
 & & &  \\[-8pt]
Site1 & O1+(90$^\circ$,31$^\circ$)	&  0.0241 	& 105.4 \\ 
Site2 & O2+(13$^\circ$,180$^\circ$)	&  0.0250 	&   90.2  \\
 & & &  \\[-8pt]
\hline\hline
	\end{tabular}
\end{table}

\begin{figure}[t]
	\centering
	\includegraphics[width=0.9\linewidth,clip]{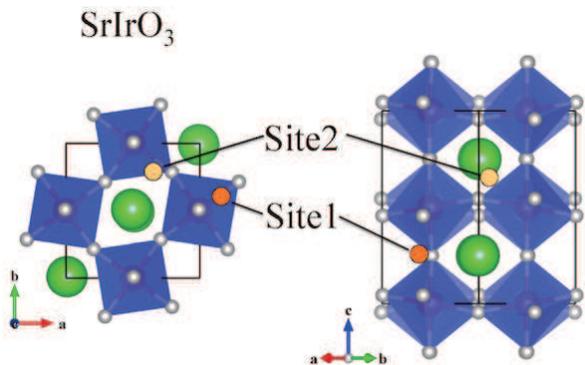}
	\caption{(Color online) Presumed muon sites in \sroi. Site1 is close to apical oxygen (O1), and Site2 is near IrO$_2$ planes (O2).} \label{musite3}
\end{figure}

\subsection{Magnetism below $T_{\rm m}$}

As noted earlier, our extensive survey on the magnetic structure vs $B_{\rm loc}$ in \sro\ and \sros\ indicated that the internal field observed below $T_{\rm m}$ cannot be explained by the alteration of the Ir$^{4+}$ isospin configuration from that established above $T_{\rm m}$ (in contrast to the suggestion in the previous report\cite{Franke:11}).  The main difficulty lies in the fact that such a scenario always leads to the modulation of $B_{\rm loc}$ at every muon site, whereas only those muons situated near the apical oxygen (O1) actually detected an increase of $B_{\rm loc}$ below $T_{\rm m}$.

Given this situation, one of the simplest hypotheses to explain such an observation would be the emergence of  ordered magnetic moments on the apical oxygen.  In general, the $d$ electron orbitals are strongly hybridized with $p$ orbitals of the ligand oxygen, which makes it plausible to have an unpaired hole on the oxygen atoms under strong electronic correlation.  The small Ir moment in the iridium compounds ($\simeq$0.3--0.4$\mu_B$) probably originates from the Ir $d$-O $p$ hybridization in addition to quantum fluctuations. Such a covalency effect has been observed in potassium chloroiridate (K$_2$IrCl$_6$, antifluorite crystal structure), where  $\sim$30\% of the transferred moment is localized on the apical Cl sites of IrCl$_6$ octahedra.\cite{Lynn:76}  Similar examples have also been found in a variety of compounds including K$_2$CuF$_4$ (Ref.\onlinecite{Hirakawa:74}), La$_{0.8}$Sr$_{0.2}$MnO$_3$ (Ref.\onlinecite{Pierre:98}), La$_5$Ca$_9$Cu$_{24}$O$_{41}$ (Ref.\onlinecite{Matsuda:98}), and Li$_2$CuO$_2$ (Ref.\onlinecite{Chung:03}). 

Moreover, the rotation of IrO$_6$ octahedra and associated changes in the Ir-O-Ir bond angle at lower temperatures\cite{Crawford:94,Li:13} may modify the $d$-$p$ hybridization, and eventually lead to oxygen polarization below $T_{\rm m}$.

\begin{figure}[t]
	\centering
	\includegraphics[width=0.9\linewidth,clip]{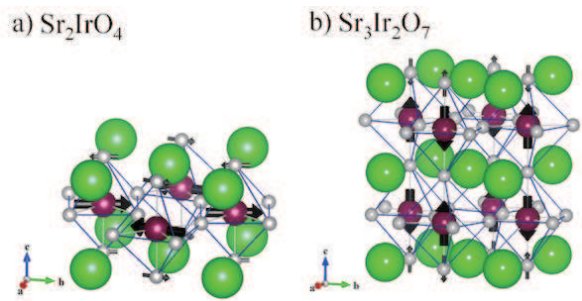}
	\caption{(Color online) Presumed structure of oxygen magnetic moments associated with apical oxygen (O1) in (a) \sro\ and (b) \sros. Oxygen moments are anti-parallel with those of iridium in each compound.} \label{O1spin}
\end{figure}

The possibility of attributing the observed change of $B_{\rm loc}$ to the spin polarization of apical oxygen (O1) was examined in a simulation of $B_{\rm loc}$ assuming a point-like magnetic dipole ${\bm \mu}_{\rm O}$ at each O1 site with its direction anti-parallel with the nearest Ir isospin moment (see Fig.~\ref{O1spin}).  $B_{\rm loc}$ was obtained by adding the dipolar fields from O1 to that from Ir (with a moment size of $0.4\mu_B$ in \sro\ and $0.3\mu_B$ in \sros).  Table \ref{tabsim} summarizes the results.  The size of ${\bm \mu}_{\rm O}$ ($=\mu_{\rm O}$) was as small as 0.02--0.03$\mu_B$, which can explain the change in $B_{\rm loc}$ at Site1 and Site1' in \sro\ while $B_{\rm loc}$ at the other muon sites was virtually unaffected (relative change of a few percent or less).   It also provides a means to map $f_1$ and $f_2$ properly to Site1 and Site1', as $B_{\rm loc}$ was greater at the Site1' than at the Site1.   The results of a simulation under a similar assumption for \sros\ suggest that $\mu_{\rm O}\simeq 0.09\mu_B$ at the O1 site sufficiently explains the magnitude of $f_4$.  A simulation assuming a ferromagnetic correlation between ${\bm \mu}_{\rm O}$ and ${\bm \mu}_{\rm Ir}$ yielded a similar result, except that $B_{\rm loc}$ at Site2/2' and Site3 tended to show a slight increase.

The origin of the difference in the temperature dependence between $f_{1,2}$ in \sro\ and $f_4$ in \sros\ below $T_{\rm m}$ is yet to be identified. It is clearly related with the origin of $f_4$ in \sro\ that emerges prior to $f_{1,2}$ with decreasing temperature. The long-range structure of oxygen moments just below $T_{\rm m}$  in \sro\ may differ from that at the lowest temperature. The magnitude of  $\mu_{\rm O}$  is another factor that may determine the effective interaction between oxygen moments. A small $\mu_{\rm O}$ in \sro\ would make it more difficult to establish a long-range order than that in \sros.

Such a difference between \sro\ and \sros\ is a strong argument against the alternative scenario that the observed behavior below $T_{\rm m}$ is due to the migration of muons from one site to another by thermal activation.  This is  because there seems to be no obvious reason for the difference in activation energy (which would be mostly determined by the muon binding energy to the apical oxygen) between these two cases.

The possibility of muon-induced local modification to the magnetic structure is also unlikely. Considering the number of implanted muons (one in the specimen at each positron decay measurement using a continuous beam),  such a local change (just around the implanted muon), would have difficulty with inducing a well-defined thermodynamical phase having a different transition temperature ($T_{\rm m}$) from that of the bulk ($T_{\rm N}$). 

\begin{table}[t]
	\caption{Summary of simulation for $B_{\rm loc}$  in \sro\ and \sros\ assuming a point-like magnetic dipolar moment ($\mu_{\rm O}$) placed on the apical oxygen with their orientation  antiparallel to Ir moment  (with Ir moment size of $0.4\mu_B$ in \sro\ and $0.3\mu_B$ in \sros, respectively). The values of $f_{i}$ are those linearly extrapolated to $T=0$.\label{tabsim}}
	\begin{tabular}{llcccccc} \hline\hline
 & Site\#  & $i$& $f_i$ (MHz) & $B_i$ (mT) & $\mu_{\rm O}$	& $B_{\rm loc}$ (mT) \\ \hline 
 & & & & & & \\[-8pt]
 & Site1& 1 & $\sim$4.5 & $\sim$33 &  & 35.7 \\ 
 \sro\ & Site1'& 2 & $\sim$3.9 & $\sim$29 & 0.025$\mu_B$ & 26.4 \\ \cline{2-5}
  & & & & & & \\[-8pt]
 &  Site2 & 3 & 8.22(4) & 60.6(3) &   & 60.1 \\ 
 & & & & & & \\[-8pt]
\hline
 & & & & & & \\[-8pt]
 & Site1 & 4 & 13.4(1) & 98.9(1) &   	&  102.0   \\  \cline{2-5}
  & & & & & & \\[-8pt]
\sros\ & Site2/2' & 2 & 0.3-0.7 & 2.2-5.2 & 0.09$\mu_B$ & 1.6-2.8 \\
 & Site3 & 3 & 9.07(3) & 66.9(2) &  & 63.1 \\
 & & & & & & \\[-8pt]\hline\hline
	\end{tabular}
\end{table}

The present model is also supported by the fact that, assuming an anitferromagnetic interaction between ${\bm \mu}_{\rm O}$ and ${\bm \mu}_{\rm Ir}$  (as might be expected for the nearly $180^\circ$ O-Ir-O bond angle), the reduction in the uniform magnetization below $T_{\rm M}\sim100$ K in \sro\ and $T_{\rm D}\sim50$ K and in \sros\ (Refs.\onlinecite{Chikara:09,Li:13}) can be qualitatively understood to be a result of its partial cancelation by oxygen polarization.

One note of interest is a recent example of the manganese compound TbMn$_2$O$_5$, where an antiferromagnetic spin polarization at the oxygen sites was observed with oxygen $K$-edge resonant x-ray scattering.\cite{Beale:10}  The compound exhibited an incommensurate antiferromagnetic order below 43 K that was associated with ferroelectric polarization because of the strong magneto-electric coupling. It also showed the development of oxygen spin polarization below 38 K, where the dielectric constant exhibited a peak, and the origin of such multiferroic behavior was attributed to a charge redistribution around the ion nucleus in accordance with the alteration of the spin configuration, where spin polarization of the oxygen is a crucial component in the relevant theoretical model.\cite{Moskvin:08}  The situation seems to be parallel with the case of \sro\ for which a similar anomaly in the dielectric constant was reported around 70-80 K, which is near $T_{\rm m}$.\cite{Chikara:09}

\section{Conclusion}

We used \msr\ to show that \sro\ and \sros\ exhibit clear signs of a two-stage magnetic order, whereas \sroi\ stays a non-magnetic metal.  Provided that the predominant muon sites are located near the oxygen atoms common to these two compounds and that the effective Ir moment size is 0.3--0.4$\mu_B$, the magnitude of $B_{\rm loc}$ in the first stage order ($T_{\rm m} \le T \le T_{\rm N}$) closely agrees with that predicted from the magnetic structures of each compound, which were recently inferred from resonant x-ray and neutron diffraction. For the secondary magnetic phase emerging below $T_{\rm m}$, we postulate that it originates from ordered magnetic moments ($\sim$0.03$\mu_B$ in \sro\ and $\sim$0.09$\mu_B$ in \sros) at the apical oxygen of the IrO$_6$ octahedron, which is induced by strong hybridization of Ir $d$-O $p$ orbitals.  This hypothesis provides a qualitative explanation for the reported anomalies in both the magnetic and dielectric properties over the relevant temperature range according to the earlier literature.  A neutron diffraction study is currently underway to confirm the oxygen magnetic moments in those compounds. 

The present results demonstrate the usefulness of the \msr\ technique in investigating local magnetism associated with oxygen, and it has the potential to be widely applicable to strongly correlated transition metal oxides owing to the fact that implanted muons tend to form a well-defined O-$\mu$ bonding state as interstitial pseudo-hydrogen atoms in oxides.  

We express our gratitude to T. Arima, S. Fujiyama, Y. Murakami, H. Nakao, H. Okabe, and K. Yamada for helpful discussion.  We also thank Hua Li for his help in calculating the electrostatic potential using VASP code and the TRIUMF staff for their technical support during the $\mu$SR experiment.  This work was partially supported by the Condensed Matter Research Center, Institute of Materials Structure Science, KEK.

\end{document}